\documentclass[11pt,a4paper]{article}
\pdfoutput=1

\usepackage{epsfig,amsfonts,amssymb,amsmath}
\usepackage{hyperref}
\usepackage{cite,upgreek}
\input epsf.sty
\usepackage{array,verbatim}
\newcolumntype{P}[1]{>{\centering\arraybackslash}p{#1}}
\newcolumntype{M}[1]{>{\centering\arraybackslash}m{#1}}

\usepackage[normalem]{ulem}
\usepackage[utf8]{inputenc}
\usepackage[T1]{fontenc}
\usepackage{lmodern, textcomp}
\usepackage[english]{babel}
\usepackage{csquotes,cancel}

\usepackage[left=1in,right=1in,top=.8in,bottom=.8in]{geometry}

\def\ZZZ{{\hbox{ Z\kern-1.6mm Z}}}
\def\RRR{{\hbox{ R\kern-2.4mm R}}}
\def\CCC{{\hbox{ C\kern-2.0mm C}}}
\def\zzz{{\hbox{z\kern-1mm z}}}

\newcommand{\f}{\frac}

\newcommand{\qeq}{{\hbox{=\kern-2.3mm ? \kern.5mm }}}
\renewcommand{\qeq}{=}

\newcommand{\ra}{\rangle}
\newcommand{\la}{\langle}

\newcommand{\bw}{\bar w}
\newcommand{\bz}{\bar z}

\newcommand{\non}{\nonumber}
\newcommand{\be}{\begin{eqnarray}}
\newcommand{\ee}{\end{eqnarray}}
\newcommand{\ben}{\begin{eqnarray}\displaystyle}
\newcommand{\een}{\end{eqnarray}}

\newcommand{\p}{\partial}
\newcommand{\sectiono}[1]{\section{#1}\setcounter{equation}{0}}

\newcommand{\lm}{\lambda}
\newcommand{\blm}{\bar{\lambda}}
\newcommand{\mc}{\mathcal}

\def\one{{\hbox{ 1\kern-.8mm l}}}
\def\zero{{\hbox{ 0\kern-1.5mm 0}}}

\newcommand{\bea}[1]{\begin{eqnarray}\label{#1} }
\newcommand{\eea}{\end{eqnarray}}

\usepackage{bm}
\usepackage[table]{xcolor}

\def\sknote#1{{\color{blue} #1}}

\allowdisplaybreaks

\begin{document}

\baselineskip 24pt

\begin{center}
{\Large \bf  One Loop Mass Renormalization of Massive States Using Pure Spinor Formalism} 

\end{center}

\vskip .6cm
\medskip

\vspace*{4.0ex}

\baselineskip=18pt

\begin{center}

{\large 
\rm  Sitender Pratap Kashyap$^{a}$, Mritunjay Verma$^{b}$ }

\end{center}

\vspace*{4.0ex}


\centerline{ \it \small $^a$ The Institute of Mathematical Sciences, IV Cross Street Road,}
\centerline{\it \small   CIT Campus, Taramani, Chennai 600113, India}

\centerline{ \it \small $^b$ Mathematical Sciences and STAG Research Centre, University of Southampton,}
\centerline{\it \small  Highfield, Southampton SO17 1BJ, UK}
\vspace*{1.0ex}
\centerline{\small E-mail:  sitenderpk@imsc.res.in, m.verma@soton.ac.uk}

\vspace*{5.0ex}

\centerline{\bf Abstract} \bigskip
As a check of the first massive integrated vertex operator in the pure spinor formalism constructed in arXiv:1802.04486, we compute the one loop 2-point function of the stable non BPS massive states in $SO(32)$ heterotic string theory. This allows us to compute the one loop renormalized mass of these states using the pure spinor formalism. Our results are in agreement with the corresponding results obtained by Sen using the RNS formalism. This provides an instance of the equivalence between the RNS and the pure spinor formalism for the massive states at loop level. 

\vfill


\vfill \eject

\baselineskip 18pt

\tableofcontents

\sectiono{Introduction } 

The string scattering amplitudes are the most important ingredients for probing the perturbative aspects of string theory. Due to this reason, the calculation of scattering amplitudes received a lot of attention since the early days of string theory. There are numerous results at tree and loop level for the massless scattering amplitudes. In contrast, the scattering of massive states has not received similar attention due to the complexities associated with these states. However, even at the perturbative level, the massive states encode a lot of interesting physics which require the knowledge of their scattering amplitudes \cite{Pius:2013sca,Pius:2014iaa,Sen:2016gqt}. Moreover, the massive states also become important at low energy scale in the case of large extra dimensions \cite{nima, Anchordoqui:2008hi
}. Some black hole states can also be thought as massive string excitations (see e.g., \cite{Sen, Duff:1994jr}) and their scattering can be analysed from the scattering of the elementary strings. Thus, the computation of the scattering amplitudes of massive string states is a crucial requirement. 

A lot of success in computing the massless string scattering has come from the development of the pure spinor approach to superstring theory \cite{Berkovits:2000fe}. The RNS formalism becomes very complicated for the loop amplitudes and the higher point functions (due to complexities such as requirement of sum over spin structures). On the other hand, the pure spinor formalism has been used very successfully to compute the massless string amplitudes which were very difficult in the RNS approach \cite{BerkovitsNekrasov,Berkovits9,Berkovits_Mafra,Berkovits10,Stanh,Gomez:2009qd,mafrathesis,Gomez:2010ad,Gomez:2013sla,Gomez:2015uha} (see \cite{Mafra:2022wml, Berkovits:2022fth} for recent reviews). However, a similar success for the massive state scatterings has been missing even in the pure spinor formalism. One reason for this in the pure spinor formalism is that the vertex operators of massive states were not known until very recently. 

\vspace*{.07in}In \cite{Berkovits1}, the first massive unintegrated vertex operator in pure spinor formalism was constructed in open string theory. However, the equations satisfied by the superfields appearing in the vertex operator were worked out only in the rest frame. Due to this, the theta expansion needed for computing the scattering amplitudes could not be worked out. This problem was solved in  \cite{theta_exp} and a systematic method to construct any higher massive integrated or unintegrated vertex operator was given in \cite{integrated} which also constructed the integrated version of the first massive vertex operator. Using these results, \cite{Chakrabarti:2018bah} showed that the pure spinor tree level amplitudes involving the massive states agree with the corresponding RNS results. This only made use of the unintegrated vertex operator. So far, there has been no computation involving the integrated massive vertex. 

\vspace*{.07in}In this paper, we initiate the computations using the first massive integrated vertex. For this, we shall consider a simple example, namely, one loop 2-point function of first massive state in the Heterotic string theory. Our motivation is two fold. First, we want to show that the loop calculations in pure spinor formalism involving the massive string states are also consistent with the RNS results. Secondly, in long term, we want to study the 2-loop mass renormalization of massive states using the pure spinor formalism. 

\vspace*{.07in}A generic string state, unless protected by some symmetry, undergoes mass renormalization. However, one can not compute the renormalised masses using the standard Polyakov formulation of string theory. The reason for this is that one needs to go off-shell for computing the renormalised masses. In a quantum field theory, if we are interested in computing the renormalised mass, we need to reorganise the perturbative expansion for 2-point Green's function in terms of 1PI and 1PR diagrams. This perturbative series can be resummed and the location of pole in this gives the renormalised mass. However, the Polyakov's approach to string perturbation theory computes a single diagram at each loop order without distinguishing between the 1PI and 1PR contributions. This situation was remedied in \cite{Pius:2013sca,Pius:2014iaa} which gave a prescription to go off-shell in the Polyakov's approach (also see \cite{Witten:2013pra}). The most efficient way to do this is to use the string field theory which neatly separates the 1PI and 1PR contributions of a given amplitude (see \cite{deLacroix:2017lif} for a review). The off-shell version of the pure spinor formalism is still not developed. Hence, we can't use an off-shell method to compute the mass renormalization using the pure spinor formalism. However, at one loop, we can compute the mass renormalization by remaining on-shell since there is no internal free propagator connecting two 1PI blobs which diverges on-shell. Hence, this computation can be performed in pure spinor as well. 

\vspace*{.07in}As mentioned above, we shall work in Heterotic string theory. The first massive states in SO(32) Heterotic string theory are stable non BPS states and they undergo mass renormalization due to loop effects \cite{Pius:2013sca, Sen:2013oza}. The mass renormalization of these states can't be computed by examining the poles in the S-matrix of the massless states since these massive states do not appear as single particle intermediate states in the scattering of the massless particles of the theory \cite{Pius:2013sca}. For this reason, one needs to compute the renormalized masses of these states directly. Another feature of these states is that they don't mix with the unphysical states when they undergo mass renormalization unlike a generic string state \cite{Pius:2013sca}. These features make it ideal for studying the mass renormalization in string theory as well as comparing the massive one loop calculations in RNS and pure spinor formalisms. One reason for focussing on pure spinor formalism for this purpose is that as in the case of massless states, the RNS formalism is, in general, very difficult to use if we want to compute the higher genus amplitudes involving the massive states to analyse such effects.   

\vspace*{.07in}The rest of the paper is organized as follows. In section \ref{sec:closedvertex}, we give the expressions of the integrated and unintegrated SO(32) Heterotic first massive closed string vertex operators, that we shall require in this work. In section \ref{sec:2loop}, we compute the one loop 2-point function of our massive states and their one loop renormalized mass and show that the results are in agreement with \cite{Sen:2013oza}. We conclude in section \ref{outlook} with a brief discussion. In the various appendices we provide additional details required in the main draft. In appendix \ref{sec2:review}, we review the Heterotic strings in the pure spinor formalism. We focus on the world-sheet theory, one loop amplitude prescription and the first massive states of the theory. Appendix \ref{appen:A} gives some details about the relation between the open and closed string vertex operators and in appendix \ref{appen:PS}, we note some useful pure spinor identities involving the massive superfields which are useful in simplifying our calculations.

\section{Massive Vertex Operators for Closed Strings}
\label{sec:closedvertex}
In this section, we construct the integrated and unintegrated vertex operators for the closed strings using the corresponding open string results. More details can be found in appendix \ref{appen:A}. Since we shall describe the supersymmetric sector of the world-sheet theory in terms of the superfields $B_{mnp},G_{mn}$ and $\Psi_{m\alpha}$, we start by writing down the equations satisfied by these superfields. As described in appendix \ref{appen:A}, the closed string vertex operators can be obtained from the open string vertex operators by rescaling the momenta as $k\rightarrow \f{k}{2}$. This is also true for the equations of motion satisfied by the superfields. Thus, the superspace equations satisfied by the superfields describing the supersymmetric sector of the theory are thus obtained from the open string result \cite{theta_exp} as 
 \be
D_{\alpha}G_{sm} = {8}i\ k^{p}(\gamma_{p(s}\Psi_{m)})_\alpha \label{D_Gmn1}
\ee
\be
D_\alpha B_{mnp}=12 (\gamma_{[mn}\Psi_{p]})_\alpha + {6}i\alpha' k^t k_{[m}(\gamma_{|t|n}\Psi_{p]})_\alpha                             \label{D_Bmnp_general}
\ee
\be
D_{\alpha}\Psi_{s\beta}= \f{1}{16} G_{sm}\gamma^{m}_{\alpha\beta}+{\f{i}{48}}k_mB_{nps}(\gamma^{mnp})_{\alpha\beta}
-{\f{i}{288}}k^mB^{npq}(\gamma_{smnpq})_{\alpha\beta} \label{D_Psi1}
\ee
with the constraints
\be
 (\gamma^{m})^{\alpha \beta}\Psi_{m\beta}=0\quad ; \quad \p^m \Psi_{m\beta}=0\quad ; \quad  \p^m B_{mnp}=0 \quad ; \quad  \p^m G_{m n}=0 \;\; \;; \;\; \eta^{mn}G_{mn}=0\label{cons_theta=0}
\ee
The mutual consistency of these equations is easy to check. For this, we take the covariant derivative of both sides of these equations and then use the identity \eqref{susy_iden} for the left hand side and covariant derivative of the superfield in the right hand side. The both sides then agree as expected.

\subsection{Unintegrated Vertex Operator}

The unintegrated vertex operator $V(z,\bar z)$ in the closed string theory satisfy \cite{Berkovits:2004px}
\be
QV(z,\bar z)=0=\bar QV(z,\bar z)\qquad,\qquad \delta V = Q\Omega+\bar Q\bar\Omega
\ee
with the condition $\bar Q\Omega=Q\bar\Omega=0$.

\vspace*{.07in}The equations for both the left and right sectors are essentially same as the ones arising in the open string case. Thus, the closed string vertex operator is essentially the tensor product of two copies of the open string operator. A more detailed analysis shows that if $V(k,z)$ denotes the open string unintegrated vertex operator with momentum $k$, then the closed string unintegrated vertex operator is given by (see appendix \ref{appen:A} for more details) 
\be
V(k,z,\bar z)= \kappa V_{{L}}\left(\f{k}{2},z\right)V_{{R}}\left(\f{k}{2},\bar z\right)e^{ik\cdot X}\label{3.14}
\ee
where $\kappa'$ is the overall normalization constant for the vertex operator.
The dependence of $V_L$ and $V_R$ upon the $X^m$ fields is only through their derivatives $\p X^m$. The fact that we need to take the tensor product of two copies of the open string vertex operator with only half of the momenta is also clear from the difference in the OPEs of the open and closed strings. As discussed in appendix \ref{seca.1}, the closed string OPEs can be obtained from the corresponding open string OPEs by the replacement $k\rightarrow \f{k}{2}$. For the type II theories, both the $V_{L}$ and $V_R$ in the right hand side of \eqref{3.14} have same functional form. However, for the heterotic strings, only the holomorphic factor corresponds to the supersymmetric open string. The anti-holomorphic factor is constructed separately as in the RNS formalism. We are interested in the heterotic string case for which $V_L$ can be written down by using the open string pure spinor result to be\footnote{In this paper, we are using the conventions for OPEs etc. of \cite{Gomez:2010ad}. The various factors of $\f{\alpha'}{2}$ in the vertex operators have been inserted as compared to the result given in \cite{theta_exp} due to this change in the conventions.  }.
\be
V_L\ =\ :\partial \theta^\beta \lambda^\alpha B_{\alpha\beta}: 
+\;\f{\alpha'}{2}\;:d_\beta\lambda^\alpha C^\beta_{\;\alpha}:+:\Pi^m\lambda^\alpha H_{m\alpha}:
+\;\f{\alpha'}{2}\;:N^{mn}\lambda^\alpha F_{\alpha mn}:                                                    \label{vertex_ansatz}
\ee
where, the superfields are only functions of $\theta^\alpha$ (since we have included the factor of $e^{ik\cdot X}$ separately) and are given in terms of the basic superfields $B_{mnp}$ and $\Psi_{m\alpha}$ to be \cite{Berkovits1}
\be 
&&H_{s\alpha}=\f{3}{7}(\gamma^{mn})_{\alpha}^{\;\;\beta}D_{\beta}B_{mns}=-72 \Psi_{s\alpha}\;, \quad C_{mnpq}={\f{i}{4}}k_{[m}B_{npq]} \;,  \non\\[.2in]
&&\hspace{.2 in}F_{\alpha mn} ={\f{i}{16}}\biggl(7k_{[m}H_{n]\alpha} + k^q(\gamma_{q[m})_\alpha^{\;\;\beta}H_{n]\beta}\biggl)
\label{2.14}
\ee
Note that in the above equations, we have replaced $k_m$ of open string solution by $\f{1}{2}k_m$. 

\vspace*{.07in}Now, for the heterotic strings, the right moving factor $V_R(\bar z)$ is same as in the RNS formalism and is given by\footnote{For the type II theories, $V_R$ has the same form as $V_L$ in equation \eqref{vertex_ansatz}.}
\be
V_{R}(\bar z)= \hat c(\bar z)\hat\rho^A(\bar z) \hat\rho^A(\bar z) \equiv \hat c(\bar z) \hat J(\bar z)\label{bar_J}
\ee
Before proceeding further, we note that the conformal weight of $e^{ik\cdot X}$ for the closed string is given by $(\f{\alpha' k^2}{4},\f{\alpha' k^2}{4})$ and the mass of the state at $n^{th}$ level is given by $k^2=-\f{4n}{\alpha'}$. This means that the conformal weight of $e^{ik\cdot X}$ at mass level $n$ is $(-n,-n)$. Demanding the conformal weight of the unintegrated vertex operator to be $(0,0)$, we find that the conformal weights of $V_L$ and $V_R$ should be $(n,0)$ and $(0,n)$ respectively. The above expressions \eqref{vertex_ansatz} and \eqref{bar_J} are consistent with this for $n=1$.

\subsection{Integrated Vertex Operator}
The integrated vertex operator $U$ can also be expressed in the factorized form 
\be
U(z,\bar z)= \kappa U_{L}(z)U_{R}(\bar z)e^{ik\cdot X}\label{1.2.17}
\ee
The expression for the left moving sector involving $U_L$ can be written down using the known RNS result and is given by
\be
U_{R}(\bar z)= \hat\rho^A(\bar z) \hat\rho^A(\bar z) = \hat J(\bar z)
\ee
The factor $U_L(z)$ can be written down by using the corresponding open string result of \cite{integrated} by rescaling $k_m\rightarrow \f{1}{2}k_m$ to be \footnote{{The various factors of $\f{\alpha'}{2}$ in the definition of $U_R$ is present  as a result of the new convention that we are following. }}
\be
\f{2}{\alpha'}\ U_L(z)&=&\ :\Pi^m\Pi^n F_{mn}:\;+\;\f{\alpha'}{2}:\Pi^m d_\alpha F_m^{\;\;\alpha}:\;+\;:\Pi^m\partial\theta^\alpha G_{m\alpha}:
\;+\;\f{\alpha'}{2}:\Pi^m N^{pq}F_{mpq}:\nonumber\\
&&+\;\;\left(\f{\alpha'}{2}\right)^2:d_\alpha d_\beta K^{\alpha\beta}:\;+\;\f{\alpha'}{2}:d_\alpha\partial\theta^\beta F^\alpha_{\;\;\beta}:\;+\;\left(\f{\alpha'}{2}\right)^2:d_\alpha N^{mn}G^\alpha_{\;\;mn}:\;
\nonumber\\
&&+\;:\partial\theta^\alpha\partial\theta^\beta H_{\alpha\beta}:\;+\;\f{\alpha'}{2}:\partial\theta^\alpha N^{mn} H_{mn\alpha}:\;+\;\;\left(\f{\alpha'}{2}\right)^2:N^{mn}N^{pq}G_{mnpq}:
\label{general_Ures}
\ee
where, the superfields appearing in \eqref{general_Ures} are given as
\be
F_{mn}&=&-\f{18}{\alpha'}G_{mn}\quad\;,\qquad F_m^{\;\alpha}\ =\ {\f{144i}{\alpha'}} (\gamma^r)^{\alpha\beta}k_r\Psi_{m\beta}
 \quad\;\;, \qquad G_{m\alpha}\ =\ -\f{432}{\alpha'}\Psi_{m\alpha} \non\\[.5cm]
 F_{mpq}&=&\f{12}{(\alpha')^2}B_{mpq}-{\f{18i}{\alpha'}}k_{[p}G_{q]m}\quad,\quad\;\; K^{\alpha \beta}\ =\ -\f{1}{(\alpha')^2}\; \gamma_{mnp}^{\alpha \beta}B^{mnp}  \non\\[.5cm]
F^{\alpha}_{\;\;\;\beta}&=&-{\f{2i}{\alpha'}} (\gamma^{mnpq})^{\alpha}_{\;\;\;\beta}k_mB_{npq}\qquad,\quad\;\;
 G^{ \alpha}_{mn}\ =\ \f{48}{(\alpha')^2}\gamma_{[m}^{ \alpha\sigma}\Psi_{n]\sigma}-{\f{48}{\alpha'}} \gamma_r^{ \alpha\sigma}k^rk_{[m}\Psi_{n]\sigma}
\non\\[.5cm]
H_{\alpha \beta}\ &=&\ \f{2}{\alpha'}\gamma^{mnp}_{\alpha \beta}B_{mnp}\quad\qquad\qquad\;\;,\quad H_{mn\alpha}\ =\ -{\f{288i}{\alpha'}}\; k_{[m}\Psi_{n]\alpha}-{\f{72i}{\alpha'}}k^q(\gamma_{q[m})_\alpha^{\;\;\sigma}\Psi_{n]\sigma}\non\\[.5cm]
G_{mnpq}&=&{\f{2i}{(\alpha')^2}}k_{[m}B_{n]pq}+{\f{2i}{(\alpha')^2}}k_{[p}B_{q]mn}+{\f{3}{\alpha'}}k_{[p}k_{[m}G_{n]q]}\label{Group_theory_Ures}
\ee
For amplitude computations, we also require the theta expansions of the superfields which are given in appendix \ref{theta_exp}. 

\section{One Loop 2-point Function and mass renormalization}
\label{sec:2loop}
In this section, we compute the one loop 2 point function of the first massive states in the heterotic string theory using the pure spinor formalism.  As reviewed in appendix \ref{loopappen}, the one loop two point function is given by
\be 
\mc{A}=
\f{\kappa^2}{2}\int d^2\tau d^2z\la \la \mc{N} B\hat B:V_L(0)V_R(0)e^{ik_1\cdot X(0)}: \;:U_L(z)U_R(\bz)e^{ik_2\cdot X(z,\bz)}:\ra\ra \label{1loop2}
\ee
where, the integrated and unintegrated vertex operators are given in the previous section, $B$ and $\hat B$ are constructed using the beltrami differentials (see equation \eqref{belhyut}), $\mc{N}$ is the regulator given in \eqref{Nregulator} and the double bracket $\la\la\cdots\ra\ra$ denotes the fact that we have gotten rid of the non-zero modes by making use of the OPEs. We denote the external momenta as $k_1$ and $k_2$ and take them to be incoming so that the on-shell momentum conservation implies
\be
k_1+k_2=0
\ee
We first consider the pure spinor part of the calculation in \eqref{1loop2}. As discussed in appendix \ref{loopappen}, all the world-sheet fields can be expanded in a complete set of basis of $\bar\p$ operator. The zero and the non zero modes in this expansion behave differently. The non zero modes of the various world-sheet fields have OPEs between them as given in section \ref{sec2:review}. However, the zero modes of different fields do not have OPEs between them. As the first step in calculation, one uses the OPEs between the non zero modes of the world-sheet fields present in the vertex operators and $b$ ghosts to eliminate them from the integrand in \eqref{1loop2}. 

\vspace*{.07in} Once we get rid of non zero modes of all the fields from the integrand using their OPEs, we shall have to perform the path integration over non zero modes with weight factor $e^{-S}$. The result of this integration is given by
\be
[\mbox{det}'{\bar\p}]^{16}[\mbox{det}'{\bar\p}]^{-11}[\mbox{det}'{\bar\p}]^{-11}[\mbox{det}'{\bar\p}]^{11}= [\mbox{det}'{\bar\p}]^{5}\label{non0mode}
\ee
where, the various factors in the left hand side come from the integrations over the $(p_\alpha,\theta^\beta)$, $(\lambda^\alpha,w_\beta)$, $(\bar\lambda^\alpha,\bar w_\beta)$ and $(r_\alpha,s^\beta)$ systems respectively. The primes in \eqref{non0mode} denote that the zero modes are excluded from the determinant. Note that the non-zero mode measures are such that the overall coefficient after integration comes out to be 1 \cite{Berkovits:2005bt}. 

\vspace*{.07in}After this, we need to perform the integration over the zero modes. Our task is simplified by the fact that for the Grassman odd fields, there are not many ways to saturate the zero modes and hence, most of the terms in the integrand do not contribute. We shall denote the zero modes of a world-sheet field $\phi$ by $\phi^1$. We start with the field $s^\alpha$. This field only appears in the regulator $\mathcal N$ and the $b$ ghost. However, in the $b$ ghost, it appears in the combination $s^\alpha\p\bar\lambda_\alpha$ and hence it can only contribute if there is $\bar w^\alpha$ in the integrand (otherwise it vanishes since it does not provide any zero mode of $\bar\lambda$). We have chosen our vertex operators to be independent of the non-minimal variables. Thus, this term will never contribute. Noting that $s^\alpha$ has 11 zero modes on torus, this means that these zero modes can only be saturated by the regulator through the factor $\mathcal{N}\rightarrow (s^1d^1)^{11}$.

\vspace*{.07in}Next, we consider the 16 zero modes of $d_{\alpha}$. The regulator $\mc{N}$ has already supplied $11$ zero modes. The remaining 5 must come from the $b$ ghost and the vertex operators. The $b$ ghost can supply either one or two zero modes of $d_\alpha$. For the two point function we are interested in, the vertex operators can supply a maximum of 3 zero modes. Thus, it follows that we must choose 2 zero modes from the $b$ ghost and 3 zero modes from the vertex operator. The 3 zero modes from the vertex operators can only come in a unique way, namely, we must pick only the $d_\alpha$ containing term in the unintegrated vertex and the $d_\alpha d_\beta$ containing term in the integrated vertex operator. Using the expressions of the vertex operators given in the previous section, the two point function can, thus, be written as
\be 
\mc{A}&=&  
\f{\kappa^2}{2}{\left(\f{\alpha'}{2}\right)^3}\int d^2\tau\;\; \biggl\la \mc{N}(y) \;B\hat B\; :d_\alpha \lambda^\beta C^{\alpha}_{\;\beta}:V_R(z)e^{ik_1\cdot X(z)} \non\\
&&\hspace*{1.5in}\int d^2w \;:d_\sigma d_\tau K^{\sigma\tau}:U_R(w)e^{ik_2\cdot X(w)} \biggl\ra\label{amp234}
\ee
Next, we compute the beltrami differential $B$. For this, we write $d_\alpha$ as
\be
d_\alpha(x) = \hat d_\alpha(x) + d^1_\alpha \omega_1(x)
\ee
The $\omega_1$ are the holomorphic 1-form given by $\omega_1=1$ on torus. The non zero mode $\hat d_\alpha$ will have OPE with the other $\hat d_\alpha$ as well as the non zero modes of $\theta^\alpha $ and $X^m$ fields. However, these OPE terms will not contribute since they will not provide the required number of zero mode $d_\alpha^1$. Thus, in the expression \eqref{amp234}, we can replace $d_\alpha$ by $d^1_\alpha$ everywhere. 
Noting that we need to pick the term containing two $d_\alpha$ from the $b$-ghost, namely
\be 
\f{\alpha'}{2}\f{(\blm \gamma^{mnp}r)(d\gamma_{mnp}d)}{192(\lm\blm)^2}
\ee
 the left moving Beltrami differential term becomes
\be
B &=& \f{1}{2\pi}\int d^2x\ b(x)\mu(x)\non\\
&=&\f{1}{2\pi}\int d^2x \f{\alpha'}{2}\f{(\blm \gamma^{mnp}r)(d^1\gamma^{mnp}d^1)}{192(\lm\blm)^2}\omega_1\omega_1\mu\non\\
&=&\f{\alpha'}{2}\f{(\blm \gamma^{mnp}r)(d^1\gamma^{mnp}d^1)}{2\pi\times 192(\lm\blm)^2}
\ee
where we used $\int d^2z w_1w_1 \mu=1$ in going to the last line.

\vspace*{.07in}Using the above results, the 2-point function can be written as 
\be 
\mc{A}&=& \left(\f{\alpha'}{2}\right)^4\f{\kappa^2}{4 \pi\times 192}\int d^2\tau \Big\la \mc{N}(y) \hat B \;\f{(\blm \gamma_{mnp}r)(d^1\gamma^{mnp}d^1)}{(\lm\blm)^2} d^1_\alpha \lambda^\beta C^{\alpha}_{\;\beta}(z)V_Re^{ik_1\cdot X(z)} \non\\
&&\times \int d^2w \;d^1_\sigma d^1_\tau K^{\sigma\tau}(w)U_Re^{ik_2\cdot X(w)} \Big\ra \label{adfg}
\ee
Next, we use the trick that $r_\alpha$ present in the $b$ ghost term in \eqref{adfg} can be replaced by the covariant derivative which acts on all the superfields in the integrand \cite{Berkovits_Mafra}. To see this, we note that $r_\alpha$ inside the integrand can be manipulated as (denoting all the other terms in the integrand by $T^\alpha$ and recalling that the regulator contains the term $e^{-r\theta}$)
\be
\left( r_\alpha e^{-r_\beta\theta^\beta} \right)T^\alpha &=&\left( \f{\p}{\p\theta^\alpha}e^{-r_\beta\theta^\beta}\right)T^\alpha\non\\
&=&-e^{-r\theta}(D_\alpha- \f{1}{2}\gamma^m_{\alpha\beta}\p_m)T^\alpha \non\\
&=&-e^{-r\theta}D_\alpha T^\alpha -e^{-r\theta} \f{i}{2}\gamma^m_{\alpha\beta}(k_1+k_2)_mT^\alpha \non\\
&=&-e^{-r\theta}D_\alpha T^\alpha
\ee
 In going to the second line, we used integration by parts and and the definition of covariant derivative. In going to the 3rd line, we used the fact that the integrand $T^\alpha$ contains the $X^m$ fields only through $e^{ik\cdot X}$. In going to the final line, we used the momentum conservation $k_1+k_2=0$.

\vspace*{.07in}We now use the zero mode measure for $s^\alpha$ given in \eqref{zeromodemeasure} to write
\be
[ds]e^{sd}&=&c_s(\lambda\bar\lambda)^{-3} T_{\alpha_1\alpha_2\alpha_3\alpha_4\alpha_5} \epsilon^{\alpha_1\cdots\alpha_5\rho_1\cdots \rho_{11}}\p^s_{\rho_1}\cdots \p^s_{\rho_{11}}e^{sd}\non\\
&=&{\left(\f{\alpha'}{2}\right)^{2}\f{(2\pi)^{11/2}}{2^611!5!}\f{Z_1^{11}}{R (\lm \bar\lm)^3}} T_{\alpha_1\alpha_2\alpha_3\alpha_4\alpha_5} \epsilon^{\alpha_1\cdots\alpha_5\beta_1\cdots \beta_{11}}d_{\beta_{1}}\cdots d_{\beta_{11}}
\ee
Using these results, the expression in \eqref{adfg} can be written as
\be 
\mc{A}&=& -{\left(\f{\alpha'}{2}\right)^6\f{\kappa^2}{4 \pi\times 192}\f{(2\pi)^{11/2}}{2^611!5!}\f{Z_1^{11}}{R }}  \int d^2\tau\ \int d^2w\int [d\lambda][d\bar\lambda][dw^1][d\bar w^1][dd^1][dr] [d\theta]\non\\
&&T_{\sigma_1\cdots\sigma_5}\epsilon^{\sigma_1\cdots\sigma_{5}\rho_1\cdots\rho_{11}}d^1_{\rho_1}\cdots d^1_{\rho_{11}} \;e^{-\lm\blm-w\bw -r\theta} \f{(\blm \gamma_{mnp}D)(d^1\gamma^{mnp}d^1)}{(\lm\blm)^{(2+3)}} \non\\
&&d^1_\alpha \lambda^\beta C^{\alpha}_{\;\beta}(z)  \;d^1_\sigma d^1_\tau K^{\sigma\tau}(w)e^{ik_1\cdot X(z)}e^{ik_2\cdot X(w)}\Bigl\la \hat BV_R(z)U_R(w)\Bigl\ra
\ee
We can now perform the integration over the $d_\alpha, w_\alpha$ and $\bar w^\alpha$ zero modes.  The measures for the zero modes in \eqref{zeromodemeasure} give at one loop (see, e.g., \cite{Gomez:2015uha})
\begin{gather}
T_{\sigma_1\cdots\sigma_5}\epsilon^{\sigma_1\cdots\sigma_{5}\rho_1\cdots\rho_{11}}\int [dd^1] d^1_{\rho_1}\cdots d^1_{\rho_{11}}(d^1\gamma^{mnp}d^1)d^1_\alpha d^1_\sigma d^1_\tau =11!\ 5!\ 96\ c_d(\lambda\gamma^{[m})_{\alpha}(\lambda\gamma^{n})_{\sigma}(\lambda\gamma^{p]})_{\tau}\non\\[.3cm]
\int [dw^1][d\bar w^1]\exp\left(-w^I_\alpha\bar w_I^\alpha\right)=
c_wc_{\bar w}(11!)^25!(2\pi)^{11}(\lm\blm)^{3}=\f{(\lm\blm)^{3}}{(2\pi)^{11}z_1^{22}} 
\end{gather}
Thus, after substituting $c_d$ and making use of the above results, the two point function becomes
\be 
\mc{A}&=& -{\left(\f{\alpha'}{2}\right)^2\f{(2\pi)^{3/2} Z_1^{5}\kappa^2}{2^8R }}\int d^2\tau\ \int d^2w\int [d\lambda][d\bar\lambda][dr]  [d\theta] e^{-\lm\blm-r\theta }   \;(\lambda\gamma^{m})_{\alpha}(\lambda\gamma^{n})_{\sigma}(\lambda\gamma^{p})_{\tau}\non\\
&&\times \f{(\blm_\rho \gamma_{mnp}^{\rho\delta}D_\delta)}{(\lm\blm)^{2}}  \lambda^\beta C^{\alpha}_{\;\beta}(z)  \; K^{\sigma\tau}(w)e^{ik_1\cdot X(z)}e^{ik_2\cdot X(w)}\Bigl\la \hat BV_R(z)U_R(w) \Bigl\ra
\ee
In principle, one can now perform the zero mode integrals over the remaining variables. However, as explained in \cite{Gomez:2015uha}, these integrals are precisely the integrals which one needs to perform at the tree level. Hence, after doing the theta expansion of the superfields, one can make use of the pure spinor superspace identities listed in \cite{Berkovits_Mafra, Stanh}. For this, one makes use of the bracket \eqref{A.67braket} described in appendix \ref{loopappen}. Thus, using the definition \eqref{A.67braket} and equations \eqref{2.14} and \eqref{Group_theory_Ures}, we can write the two point function in the form 
\be 
\mc{A}&=& {i\left(\f{\alpha'}{2}\right)\f{(2\pi)^{3/2} Z_1^{5}\kappa^2}{2^{11}R }}\int d^2\tau \int d^2w\ \Bigl\la \hat B V_R(z)U_R(w) \Bigl\ra \non\\
&& \;\Bigl\la (\gamma^{abcd})^\alpha_\beta \gamma_{qrs}^{\sigma\tau} (k_1)_a(\lambda\gamma^{m})_{\alpha}(\lambda\gamma^{n})_{\sigma}(\lambda\gamma^{p})_{\tau}\lambda^\beta(\blm_\rho \gamma_{mnp}^{\rho\delta}D_\delta)   B^1_{bcd}(z)  \; B_2^{qrs}(w)e^{ik_1\cdot X(z)}e^{ik_2\cdot X(w)} \Bigl\ra_{(1,1)}\non\\ \label{t2point}
\ee
We now simplify the pure spinor part of the correlator as follows
\be
\la f\ra_{(1,1)}&\equiv&\Bigl\la(\gamma^{abcd})^\alpha_\beta \gamma_{qrs}^{\sigma\tau} (k_1)_a(\lambda\gamma^{m})_{\alpha}(\lambda\gamma^{n})_{\sigma}(\lambda\gamma^{p})_{\tau}(\blm_\rho \gamma_{mnp}^{\rho\delta}D_\delta) \lambda^\beta B^1_{bcd}  \; B_2^{qrs}\Bigl\ra_{(1,1)}\non\\
&=&\Bigl\la \lambda^{\alpha_1}\lambda^{\alpha_2}\lambda^{\alpha_3}\lambda^{\alpha_4}\bar\lambda_{\beta_1}\Bigl[(k_1)_a\gamma^{mabcd}_{\alpha_1\alpha_4}\gamma^{nqrsp}_{\alpha_2\alpha_3} \gamma_{mnp}^{\beta_1\delta}D_\delta (B^1_{bcd}  \; B_2^{qrs})\Bigl]\Bigl\ra_{(1,1)}\non\\
&\equiv& \Bigl\la\lambda^{\alpha_1}\lambda^{\alpha_2}\lambda^{\alpha_3}\lambda^{\alpha_4}\bar\lambda_{\beta_1}f_{\alpha_1\alpha_2\alpha_3\alpha_4}^{\beta_1}(\theta)\Bigl\ra_{(1,1)}\non\\
&=& \Bigl\la672 \lambda^{\sigma_1}\lambda^{\sigma_2}\lambda^{\sigma_3}\mathcal{T}^{\alpha_1\alpha_2\alpha_3\alpha_{4}}_{\sigma_1\sigma_2\sigma_3\beta_{1}}f^{\beta_1}_{\alpha_1\cdots\alpha_{4}}(\theta)\Bigl\ra_{(2,1)}\label{4.61}
\ee
In going from 3rd to 4th line, we made use of the identity \eqref{theorem1} and the tensor $\mathcal{T}$ is defined by
\be
\mathcal{T}^{\alpha_1\alpha_2\alpha_3\alpha_4}_{\sigma_1\sigma_2\sigma_{3}\sigma_4}= \f{1}{2772}\left[ \delta^{(\alpha_1}_{\sigma_1}\cdots  \delta^{\alpha_4)}_{\sigma_4} -\f{1}{4}\delta^{(\alpha_1}_{(\sigma_1}\delta^{\alpha_2}_{\sigma_2}\gamma_z^{\alpha_3\alpha_4)}\gamma^z_{\sigma_3\sigma_4)}+\f{1}{160} \gamma_z^{(\alpha_1\alpha_2}\gamma^z_{(\sigma_1\sigma_2}\gamma_x^{\alpha_3\alpha_4)}\gamma^x_{\sigma_3\sigma_4)} \right]
\ee
By making use of the pure spinor superspace identities and the equation of motion \eqref{D_Bmnp_general}, the quantity in \eqref{4.61} can be evaluated with the help of Cadabra \cite{0608005,0701238} or Mathematica package GAMMA \cite{Gamma} to be 
\be
\la{f}\ra_{(1,1)}&=& - {4608}\, \Bigl\la{B^2}_{a b c} ({k_1})_{d} (\lambda {\Psi^1}_{e}) (\lambda {\gamma}_{a b c d e} \lambda) \Bigl\ra_{(2,1)}+ {6912}\, \Bigl\la{B^1}_{a b c} {(k_1)}_{d} (\lambda {\Psi^2}_{e}) (\lambda {\gamma}_{a b c d e} \lambda) \Bigl\ra_{(2,1)}\non\\
&=& {-\f{18i}{\alpha'}}N_{(2,1)}e_{mn}e^{mn}
\label{fhatfinal}
\ee
Note that the correlator $\la{f}\ra$ does not have any $\tau$ dependence coming from the OPEs since we did not need to take any OPEs in our calculation. If we consider one loop higher point amplitudes or higher genus amplitudes, then there would be more ways to saturate the $d_\alpha$ zero modes and the modular dependence will arise from the OPEs. 

\vspace*{.07in}Using the above result \eqref{fhatfinal} and equation \eqref{normpg}, the two point function \eqref{t2point} thus becomes
\be 
\mc{A}&=&162(A_1)^{-5/2} \pi^4 Z_1^5 \alpha'^2e_{mn}e^{mn}\int d^2\tau\ \int d^2w\ \Bigl\la \hat B V_R(\bar z)U_R(\bar w) \Bigl\ra \Bigl\la e^{ik_1\cdot X(z)}e^{ik_2\cdot X(w)} \Bigl\ra\label{matchpoint}
\ee
The first correlator in the second line of \eqref{matchpoint} essentially factorizes in terms of the correlator involving the ghost field of the non supersymmetric sector, namely $\la \hat b\hat c\ra$, and a correlator involving the SO(32) fields $\hat\rho^A$. These and the last correlator of \eqref{matchpoint} are essentially same as the ones which also arise in the heterotic string in the RNS formalism. Thus, we can use the results of \cite{Sen:2013oza, Atick:1986ns, Lust }
\be
\la \hat b\hat c \ra &=&\bigl[\overline {\eta(\tau)}\bigl]^2 \label{bccorre}\\
\la\hat J(\bar z)\hat J(\bar w) \ra&=&\f{1}{2}\bigl(\overline {\eta(\tau)}\bigl)^{-16}\left(\f{\overline{\vartheta'_1(0,\tau)}}{\overline{\vartheta_1(w-z)}}\right)^{4}\sum_{\nu}   \overline{\vartheta_\nu\left((w-z)/2\right)}^{16}\label{JJcorre}\\
 \Bigl\la e^{ik_1\cdot X(z)}e^{ik_2\cdot X(w)} \Bigl\ra&=& (2\pi)^{10} \delta^{10}\left(k_1+k_2 \right)(A_1)^{5}\left(-2\pi^2\alpha'\mbox{det}'\p\bar\p\right)^{-5}  \left|\f{\vartheta_1(z-w,\tau)}{\vartheta'_1(0,\tau)}\right|^4\non\\
 &&\exp\left[-4\pi\f{(Im(z-w))^2}{\tau_2}    \right]\label{bosonic_part}
\ee
The correlators in \eqref{bccorre} and \eqref{JJcorre} are given only upto the overall numerical factor (see footnote \ref{foot4}). Using these results, we can now complete the calculation of 2-point function. However, it is instructive to first count the $\tau$ dependence hidden in the normalization factors. The $\tau$ dependence comes from the factor of $Z_g$ and $A_g$ present in the pure spinor path integral measures and the zero mode integral of $X^m$ fields. The $Z_g$ appears in the measure of conformal weight one fields and its total power at one loop is $Z_1^{5}$ due to the factor $c_d c_w c_{\bar w} c_s $. This contributes $\tau_2^{-5/2}$. The $A_g$ appears in the measure of conformal weight zero fields with the factor $A_1^{-5/2}$ and due to zero mode integral of $X^m$ fields with the factor $A_1^{5}$. These contribute $(\tau_2)^{5/2}$. Thus, the $\tau$ dependence arising from the measures and zero mode integrals mutually cancel. 

\vspace*{.07in}Thus, using \eqref{coeffc}, \eqref{periodmat}, \eqref{non0mode}, \eqref{bosonic_part}, the theta function identity $\vartheta'_1(0,\tau)=-2\pi\eta^3(\tau)$ and noting that the determinant of the operator $\p$ can be expressed in terms of eta function as $\mbox{det}'(\p)=\tau_2\overline{\eta(\tau)}^2$, we finally obtain
\be 
\mc{A}&=& \f{ N}{\alpha'}e_{mn}e^{mn}\int \f{d^2\tau}{\tau_2^{5}} \int d^2w\ \left(\overline{\eta(\tau)}\right)^{-18}\eta^{-6}(\tau)\left(\f{{\vartheta_1(w-z,\tau)}}{\overline{\vartheta_1(w-z)}}\right)^{2}\sum_{\nu}  \left\{ \overline{\vartheta_\nu\left((w-z)/2\right)}^{16}\right\}\non\\
&&\hspace*{.4in}  \exp\left[-4\pi\f{(Im(z-w))^2}{\tau_2}    \right]\label{matchpoint1a}
\ee
The $N$ is the overall numerical factor. This is undetermined since the normalization of $\la  b c \ra$ correlator in the non supersymmetric world-sheet sector is not completely fixed.\footnote{\label{foot4}As discussed in \cite{DHoker:2005jhf}, the normalization of the ghost correlators in RNS is still an unsolved problem and there is no concensus even on the normalization of the simple correlators such as $\la b c \ra$ \cite{jor, soul1, soul, R1
}.} Apart from $N$, the expression matches with the corresponding expression (4.17) in \cite{Sen:2013oza}. The  \cite{Sen:2013oza} fixed the overall numerical factor by comparing the above result with the expected result in the low energy effective theory. Since the functional dependence coming from the integral is same in two cases, the numerical factor is guaranteed to be identical in RNS and pure spinor. 

\vspace*{.07in}It would be interesting to compare the numerical factors in the RNS and pure spinor. In the supersymmetric side treated using the pure spinor, we have kept track of the numerical factors. However, in the calculation using the RNS formalism in \cite{Sen:2013oza}, this is not completely fixed due to the ghost correlators. Equating the RNS and pure spinor results, the correlators coming from the bosonic world-sheet side cancel each other and we get a prediction for the numerical constant in the product of two ghost correlators $\la \hat b\hat c \ra$ and $\la \hat \beta\hat \gamma \ra$ in the RNS formalism.   

\vspace*{.07in}The one loop renormalized mass of the heterotic massive states can be expressed in terms of the two point function computed above. The shift in the mass due to the one loop effect is given by \cite{Sen:2013oza}
\be 
\delta M=M K_{w} g^2
\ee
where, $M$ is the tree level mass, $g^2$ is the string coupling constant and $K_w$ is the integral expression in \eqref{matchpoint1a} including the numerical constant for $N=-1/(64\pi)$ \cite{Sen:2013oza}. 

\section{Discussions} \label{outlook} 

In this paper, we have computed the one loop 2-point function involving the first massive string states in the SO(32) Heterotic string theory using the pure spinor formalism. This allows us to compute the mass renormalization of these states in the pure spinor formalism. Our results are consistent with the results obtained in \cite{Sen:2013oza}. This shows that loop amplitudes involving the massive states in the pure spinor formalism also agree with the corresponding RNS results. 

\vspace*{.07in}An important feature of the pure spinor calculation is that we didn't need to sum over the spin structure unlike RNS calculation. In \cite{Sen:2013oza}, same calculation using the RNS formalism required the use of Riemann identity to simplify the expressions. In the pure spinor calculation, we directly got the simplified expression without making use of the Riemann identity. This demonstrates a generic feature of the pure spinor formalism that it naturally avoids some of the mathematical complications present in RNS formalism. However, we still encounter some complications of RNS formalism in heterotic theory. We needed to use the expression of determinant of the world-sheet operator $\p$. In the pure spinor calculations in type II theories, the various determinant factors coming from left and right moving sectors cancel each other avoiding the necessity to compute the determinants \cite{Berkovits:2005bt}. This is a simplifying feature over RNS formalism. However, for the heterotic theories, this feature is absent since the left moving sector is not described by pure spinors. 

\vspace*{.07in}Since the pure spinor formalism provides a more efficient way to compute the higher genus amplitudes, it will be promising for the calculation of higher genus mass corrections to first massive non BPS heterotic states. The 2-loop correlator can be efficiently computed using the pure spinor formalism. The divergences will appear when we try to perform the integration over the moduli space. This happens due to Riemann surfaces developing long handles in some corners of moduli space. One needs to separate these contributions for calculating the renormalized mass. As mentioned in the introduction, the proper way to compute the higher genus mass corrections is to go off-shell. We don't have the pure spinor version of the string field theory at the moment. Part of this is related to the issue of pure spinor $b$ ghost and its divergences associated with the poles in $\lambda\bar\lambda$. However, the issue of $\lambda\bar\lambda$ pole is not problematic at the level of two loop we are interested in. Further, even though we don't have a working string field theory with pure spinors \footnote{See \cite{Berkovits:2005bt} for an attempt in this direction. }, we can try to go off-shell in the world-sheet approach as described in \cite{Pius:2013sca,Pius:2014iaa}. This will require, e.g., making the vertex operators dependent on the local coordinates. In the pure spinor formalism, this might also involve relaxing the pure spinor constraint. We hope to report on it in future \cite{2loop}. 

\bigskip

\bigskip

\noindent{\bf Acknowledgments:}  
We thank Subhroneel Chakrabarti for collaboration during initial stages of this work. We are also thankful to Carlos Mafra and Luis Alberto Ypanaque Rocha for discussion and Subhroneel Chakrabarti, Carlos Mafra, Rafaelle Marotta and Ashoke Sen for comments on an earlier version of this draft. SK is thankful to IOP, Bhubaneshwar and MV is thankful to INFN, Napoli where part of this work were done. The work of MV is supported in part by the STFC consolidated grant ST/T000775/1 “New Frontiers in Particle Physics, Cosmology and Gravity”.

\appendix

\section{$SO(32)$ Heterotic String in Pure Spinor Formalism}
\label{sec2:review}
In this appendix, we briefly review the results regarding the Heterotic strings in the pure spinor formalism and its first massive states. For more details on this section, see e.g., \cite{BerkovitsNekrasov,Gomez:2009qd,Gomez:2010ad,witten}. We shall follow the conventions of \cite{Gomez:2010ad} about the space-time dimensions and the path integral measures of the various world-sheet fields. 

\subsection{World-sheet Theory}
\label{seca.1}
 In the heterotic closed string theory, the world-sheet supersymmetry is kept only in the left moving sector. The right moving sector has no supersymmetry. Since the left moving sector has supersymmetry, it can be described by the pure spinor formalism. On the other hand, the right moving sector is described by an internal CFT with central charge 16 as in RNS formalism. 

\vspace*{.07in}We shall be making use of the non-minimal version of the pure spinor formalism for the loop computations \cite{Berkovits:2005bt}. The world-sheet action describing the heterotic string theory in the non-minimal pure spinor formalism is given (in the conventions followed in \cite{Gomez:2010ad} )
\be 
S &=& \f{1}{2\pi\alpha'}\int d^2z\ \Bigl(\p X^m \bar\p X_m +\alpha' p_\alpha\bar\p\theta^\alpha-\alpha'w_\alpha\bar\p\lambda^\alpha-\alpha'\bar w^\alpha\p\bar\lambda_\alpha+\alpha's^\alpha\bar\p r_\alpha\Bigl)\non\\[0.1in]&&\hspace{0.5in} +\;\;\f{1}{4\pi}\int d^2z\Bigl(  \hat\rho^A\p\hat\rho^A + 2\hat b\p\hat c\Bigl) \ \label{2.1qw}
\ee
In the supersymmetric sector, the fields $p_\alpha, \theta^\alpha, r_\alpha$ and $s^\alpha$ are fermionic in nature while the rest of the fields are bosonic in nature. The fields $w_\alpha,\bar w^\alpha,p_\alpha$ and $s^\alpha$ have conformal weight one while $\lambda^\alpha, \bar\lambda_\alpha,\theta^\alpha$ and $r_\alpha$ have conformal weight zero. For the purpose of writing down the path integral measures, it is instructive to note the target space length dimensions of the pure spinor world-sheet fields. The length dimensions of various quantities appearing in the action are 
\be
 [\theta^\alpha]=[\lambda^\alpha]=[\bar w^\alpha]=[s^\alpha]={\f{1}{2}}\;,\;
 [p_\alpha]=[w_\alpha]=[{\bar\lambda}_\alpha]=[r_\alpha]={-\f{1}{2}}\;,\;[\alpha']=2, \;\;[X^m]=1
\ee
On the Riemann surface, an arbitrary conformal weight +1 field $\phi$ can be expanded as a linear combination of the eigenfunctions of the operator $\bar\p$, i.e.,
\be
\phi = \sum_i \phi^i f_i(z,\bar z)\qquad,\qquad \bar\p f_i(z.\bar z)= \gamma_if_i(z,\bar z)
\ee
Now, it is a result from the theory of Riemann surfaces that
the object of conformal weight  +1 have $g$ zero modes on a
genus $g$ Riemann surface satisfying $\bar\p f_i=0$. Separating these zero modes, we can thus write 
\be
\phi(z) = \hat\phi(z) + \sum_{I=1}^g\phi^I\omega_I(z)
\ee
where, $\omega_I(z)$ are the $g$ holomorphic one-forms satisfying $\bar\p \omega_I(z,\bar z)=0$. The $\hat\phi$ have no zero modes. These modes satisfy
\be
\int_ {a_I} \omega_J=\delta_{IJ}\qquad,\qquad \int_{a_I}dz\ \hat\phi_I(z) =0
\ee
The objects of conformal weight 0 have one zero mode on
every genus $g$ Riemann surface. Hence, the number of zero modes of the various pure spinor world-sheet fields on a genus $g$ Riemann surface can be tabulated as
\begin{center}
\begin{tabular*}{10cm}{llllllllll}
$X^m$  & $\lambda^\alpha$ & $w_\alpha$ & $\blm_\alpha$ & $\bw^\alpha$ &$ \theta^\alpha$& $p_\alpha$& $r_\alpha$ & $s^\alpha$\\
10 &  11 & 11g & 11 & 11g & 16 & 16g & 11 & 11g
\end{tabular*}
\end{center}
Since, we do not have fundamental world-sheet fields with higher conformal weights, this is all we need. At one loop, $g=1$, so that all the fields have one zero mode.

\vspace*{.07in}As in open strings, it is convenient to define the SUSY invariant combinations
\be
d_\alpha&=&p_\alpha-\f{1}{\alpha'}\gamma^m_{\;\;\alpha\beta}\theta^\beta\partial X_m-\f{1}{4\alpha'}\gamma^m_{\alpha\beta}\gamma_{m\sigma\delta}\theta^\beta\theta^{\sigma}\partial\theta^\delta
\non\\[.3cm]\Pi^m&=&\partial X^m+\f{1}{2}\gamma^m_{\alpha\beta}\theta^\alpha\partial\theta^\beta \label{susy_momenta}
\ee
The various OPEs are given by
\be
d_\alpha(z)d_\beta(w)=-\f{2}{\alpha'}\f{\gamma^m_{\alpha\beta}}{(z-w)}\Pi_m(w)+\cdots
\quad,\qquad
d_\alpha(z)\Pi^m(w)=\f{\gamma^m_{\alpha\beta}}{(z-w)}\partial\theta^\beta(w)+\cdots\non
\ee
\be
d_\alpha(z)V(w)=\f{D_\alpha V(w)}{(z-w)} +\cdots \quad,\qquad
\Pi^m(z)V(w)=-{\f{\alpha'}{2}}\f{\partial^mV(w)}{(z-w)}+\cdots\non
\ee
\be
 \Pi^m(z)\Pi^n(w)=-\f{\alpha'\eta^{mn}}{2(z-w)^2}+\cdots\quad,\qquad N^{mn}(z)\lambda^\alpha(w)=\f{(\gamma^{mn})^\alpha_{\;\;\beta}}{2(z-w)}\ \lambda^\beta(w)+\cdots
\non
\ee
\be
J(z)J(w)=-\f{4}{(z-w)^2}+\cdots\quad,\qquad  J(z)\lambda^\alpha(w)=\f{1}{(z-w)}\lambda^\alpha(w)+\cdots\non
\ee
\be
N^{mn}(z)N^{pq}(w)=-\f{6}{(z-w)^2}\eta^{m[q}\eta^{p]n}-\f{2}{(z-w)}\Bigl(\eta^{p[n}N^{m]q}-\eta^{q[n}N^{m]p}\Bigl)+\cdots\label{OPE_eq}
\ee

In the above OPEs, $\p_m$ is the derivative with respect to the spacetime coordinate $X^m$ and $\p$ is the derivative with respect to the world-sheet coordinate. The $V(w,\bar w)$ denotes an arbitrary superfield which depends upon the world-sheet fields $X^m$ through the plane wave factor $e^{ik\cdot X}$. The $D_\alpha$ is the supercovariant derivative whose definition is taken to be different from that for the open string case 
 \be D_\alpha\equiv\p_\alpha+ {\f{1}{2}}\gamma^m_{\alpha\beta}\theta^\beta\p_m\label{susy_derivative}
\ee
This supercovariant derivative satisfies the identity
\be
\lbrace D_\alpha, D_\beta\rbrace=  (\gamma^m)_{\alpha \beta} \partial_m \quad\implies\qquad {\f{1}{8}}(\gamma_m)^{\alpha\beta}D_\alpha D_\beta = \p_m\label{susy_iden}
\ee
The above OPEs and the results in \eqref{susy_derivative} and \eqref{susy_iden} can be derived from the corresponding open string results by making the rescaling $k^m\rightarrow \f{1}{2}k^m$ or equivalently $\p_m\rightarrow \f{1}{2}\p_m$ (see \cite{integrated} for the corresponding results for open strings in our conventions). In fact, this trick is far more general than just obtaining the OPEs. E.g., for a holomorphic function, using chain rule, we have
\be
\p f(z) = \p X^m \p_m f \ +\ \p\theta^\alpha \p_\alpha f \ =\ \Pi^m\p_m f \ +\ \p\theta^\alpha D_\alpha f  \label{1.2.15}
\ee 
where, in going to the second equality, we have made use of equations \eqref{susy_momenta} and \eqref{susy_derivative}. The corresponding open string result which was derived in \cite{integrated} includes a factor of $2$ in front of the term containing $\Pi^m\p_m$. This shows that the above result \eqref{1.2.15} can be obtained from the corresponding open string result by rescaling $\p_m\rightarrow \f{1}{2}\p_m$. 

\vspace*{.07in} For the amplitude computations, one introduces a composite $b$ ghost in the supersymmetric pure spinor sector which is given by \cite{BerkovitsNekrasov}
\be
b &=& s^\alpha \p\bar\lambda_\alpha +\f{2\Pi^m(\bar\lambda\gamma_m d)-N_{mn}(\bar\lambda\gamma^{mn}\p\theta)-J_\lambda (\bar\lm \p \theta) - (\bar\lambda\p^2\theta)}{4(\bar\lambda\lambda)}\non\\[.2cm]
&&+\f{(\bar\lambda\gamma^{mnp}r)(\left(\f{\alpha'}{2}\right)^{}d\gamma_{mnp}d+24N_{mn}\Pi_p)}{192(\bar\lambda\lambda)^2}-\left(\f{\alpha'}{2}\right)\f{(r\gamma_{mnp}r)(\bar\lambda\gamma_{m}d)N_{np}}{16(\bar\lambda\lambda)^3}\non\\[.2cm]
&&+\left(\f{\alpha'}{2}\right)\f{(r\gamma_{mnp}r)(\bar\lambda\gamma^{pqr}r)N^{mn}N_{qr}}{128(\bar\lambda\lambda)^4}\label{bghost}\non
\ee

Finally, the BRST operators in the left and right moving sectors are given by 
\be
Q&=&\oint \f{dz}{2\pi i}\ \Bigl(\lambda^\alpha d_\alpha+\bar w^\alpha r_\alpha\Bigl)
\non\\
\bar Q &=& \oint \f{d\bar z}{2\pi i}\ \biggl[\hat c(\bar z)\hat T(\bar z) +\hat b(\bar z)\hat c(\bar z)\bar\p \hat c(\bar z)\biggl]
\ee
where, the right moving component of the matter stress energy tensor is given by
\be
\hat T= -\f{1}{\alpha'}\bar\p X^m \bar\p X_m -\f{1}{2}\hat \rho^A\bar\p \hat\rho^A
\ee
The total BRST operator is $Q+\bar Q$.

\subsection{Loop Amplitude Prescription}
\label{loopappen}
In this appendix, we briefly review the loop amplitude prescription in the pure spinor heterotic theory following \cite{BerkovitsNekrasov, Gomez:2010ad, Gomez:2015uha, Berkovits:2005bt}. The multiloop $n$-point closed string amplitudes are given by \cite{Berkovits:2005bt}
\be
\mc{A}^{(0)}_{n}&=&\int\prod_{j=4}^{n}d^2z_i
\la \la\mc{N}^{(0)} \;V^{1}(0)V^{2}(1)V^{3}(\infty)U^{i}(z_i,\bz_i)\ra \ra  \\
\mc{A}^{(1)}_{n}&=&S_{1}\int_{\mathcal M_1} \;d^2\tau \int\prod_{j=i}^{n}d^2z_i
\la \la\mc{N}^{(1)} \;B\hat BV^{1}(0)U^{2}(z_i,\bz_i)\ra \ra \label{1loopasd}  \\
\mc{A}^{(g)}_{n}&=&S_{g} e^{(2g-2)\lm }\;\int_{\mathcal M_g} \;\prod_{i=1}^{3g-3}\;d^2\tau_i \int\prod_{j=1}^{n}d^2z_j
\la \la\mc{N}^{(g)} \;B_j\hat B_j\;U^{j}(z_j,\bz_i))\ra \ra \quad, \;\; g\ge 2 \non
\ee
In the above expressions, $S_g$ are the symmetry factors for a genus $g$ surface. To be precise $S_1=1/2$ \cite{DHoker_Phong1,Sakai_Tanii}, $S_{2}=1/2$\cite{Bershadsky_ooguri} and $S_g=1$ for $g>2$
. Also, $\la\la\cdots \ra\ra$ stands for the correlator in which we have gotten rid of the non-zero modes by the OPE computation. $B$ and $\hat B$ involve the beltrami differential 
\be
B=\f{1}{2\pi}\int d^2w \mu(w)b(w)\qquad,\qquad \hat B=\f{1}{2\pi}\int d^2w \hat\mu(\bar w)\hat b(\bar w) \label{belhyut}
\ee
where $b(w)$ is the composite pure spinor \textquotedblleft $b$-ghost\textquotedblright defined in \eqref{bghost} and $\hat b(\bar w)$ is the elemenatary \textquotedblleft$b$-ghost\textquotedblright given in \eqref{2.1qw}. The $\mu$ and $\hat\mu$ are the beltrami differentials given on the torus by $\mu=\hat\mu=1/2\tau_2$. $\mc{M}_1$ denotes the fundamental domain of the moduli space of torus.

\vspace*{.06in}On the torus, there is only one globally defined analytic vector field (hence one conformal killing vector) which allows us to fix the position of one of the vertex operator in \eqref{1loopasd}. The regulator $\mathcal{N}$ is needed to regularize the zero mode integrals of the pure spinor world-sheet fields. To see what kind of terms are allowed for this purpose, one uses the fact that the expression inside the correlator apart from $\mathcal{N}$ is BRST invariant. We don't want to break the BRST invariance. Hence, the allowed form of the regulator should take the form \cite{BerkovitsNekrasov}
\be
\mathcal{N}(y)= e^{\{Q,\chi(y)\}}=1+\cdots
\ee
where, $\{Q,\chi(y)\}=\oint dz (\lambda^\alpha d_\alpha+\bar w^\alpha r_\alpha)\chi(y)$ and $\cdots$ terms are BRST invariant. A convenient choice for $\chi$ which works upto two loop is given by \cite{mafrathesis}
\be
\chi=-(\bar\lambda\theta)-(w^Is^I)\qquad,\qquad I= 1,\cdots, g
\ee
which gives
\be
\mathcal{N}(y)= \exp\left[ -(\lambda\bar\lambda) -(r\theta) -(w^I\bar w^I)+(s^Id^I)    \right]\label{Nregulator}
\ee
 The integration over the non zero modes of pure spinor world-sheet fields in the correlators is done using the OPEs given in the previous subsection. On the other hand, the integration over the zero modes is performed using the following measure
\be
\int [d\lambda][d\bar\lambda][dw^I][d\bar w^I][dd^I][dr][ds][d\theta]
\ee
where \cite{Gomez:2010ad},
\be
[d\lambda]T_{\alpha_1\alpha_2\alpha_3\alpha_4\alpha_5}& =& c_\lambda \epsilon_{\alpha_1\cdots\alpha_5\rho_1\cdots \rho_{11}}d\lambda^{\rho_1}\cdots d\lambda^{\rho_{11}}\non\\[.2cm]
[d\bar\lambda]\bar T^{\alpha_1\alpha_2\alpha_3\alpha_4\alpha_5}& =& c_{\bar\lambda} \epsilon^{\alpha_1\cdots\alpha_5\rho_1\cdots \rho_{11}}d\bar\lambda_{\rho_1}\cdots d\bar\lambda_{\rho_{11}}\non\\[.2cm]
[d\bar\omega] T_{\alpha_1\alpha_2\alpha_3\alpha_4\alpha_5}& =& c_{\bar\omega}(\lambda\bar\lambda)^3 \epsilon_{\alpha_1\cdots\alpha_5\rho_1\cdots \rho_{11}}d\bar\omega^{\rho_1}\cdots d\bar\omega^{\rho_{11}}\non\\[.2cm]
[d\omega]& =& c_\omega T_{\alpha_1\alpha_2\alpha_3\alpha_4\alpha_5} \epsilon^{\alpha_1\cdots\alpha_5\rho_1\cdots \rho_{11}}d\omega_{\rho_1}\cdots d\omega_{\rho_{11}}\non\\[.2cm]
[dr]& =& c_r\bar T^{\alpha_1\alpha_2\alpha_3\alpha_4\alpha_5} \epsilon_{\alpha_1\cdots\alpha_5\rho_1\cdots \rho_{11}}\p_r^{\rho_1}\cdots \p_r^{\rho_{11}}\non\\[.2cm]
[ds]& =& \f{c_s}{(\lambda\bar\lambda)^3} T_{\alpha_1\alpha_2\alpha_3\alpha_4\alpha_5} \epsilon^{\alpha_1\cdots\alpha_5\rho_1\cdots \rho_{11}}\p^{s}_{\rho_1}\cdots \p^{s}_{\rho_{11}}\non\\[.2cm]
[d\theta] &=& c_\theta d^{16}\theta\qquad,\quad  [dd^I] =c_dd^{16}d^I
\label{zeromodemeasure}
\ee
where, the various coefficients are given by 
\be
c_\lambda &=&\left(\f{\alpha'}{2}\right)^{-2}\f{1}{11!}\left(\f{A_g}{4\pi^2}\right)^{11/2}\qquad,\quad c_{\bar{\lambda}} =\left(\f{\alpha'}{2}\right)^{2}\f{2^6}{11!}\left(\f{A_g}{4\pi^2}\right)^{11/2}\non\\[.2cm]
c_w &=&\left(\f{\alpha'}{2}\right)^{2}\f{(4\pi^2)^{-11/2}}{11!5!}\f{1}{Z_g^{11/g}}\qquad,\quad c_{\bar{w}} =\left(\f{\alpha'}{2}\right)^{-2}\f{(4\pi^2)^{-11/2}}{11!}\f{(\lm \bar\lm)^3}{Z_g^{11/g}}\non\\[.3cm]
c_r &=&\left(\f{\alpha'}{2}\right)^{-2}\f{R}{11!5!}\left(\f{2\pi}{A_g}\right)^{11/2}\qquad,\quad c_{s} =\left(\f{\alpha'}{2}\right)^{2}\f{(2\pi)^{11/2}}{2^611!5!}\f{Z_g^{11/g}}{R (\lm \bar\lm)^3}\non\\[.3cm]
c_\theta &=&\left(\f{\alpha'}{2}\right)^{4}\left(\f{2\pi}{A_g}\right)^{16/2}\qquad,\quad c_{d} =\left(\f{\alpha'}{2}\right)^{-4}(2\pi)^{16/2}Z_g^{16/g}\label{coeffc}
\ee
The parameter $R$ in the above coefficients is fixed, by demanding the tree amplitudes to be same in the RNS and pure spinor calculations, to be $R^2=\f{\sqrt{2}}{2^{16}\pi}$ \cite{Gomez:2010ad} and the quantities $A_g$ and $Z_g$ are given by (denoting the period matrix by $\Omega_{IJ}$)
\be
A_g=\int d^2z\sqrt{g}\qquad,\qquad Z_g = \f{1}{\sqrt{\mbox{det}(2\mbox{Im}(\Omega_{IJ}))}}\label{periodmat}
\ee
The advantage of including these factors in the measure is that the norm of the zero mode basis becomes independent of the modular parameters \cite{Gomez:2010ad} (See \cite{Gomez:2009qd} for calculation in which measure has been kept independent of the modular parameters)

The tensors $T$ and $\bar T$ appearing in the measures are defined as 
\be
T_{\alpha_1\alpha_2\alpha_3\alpha_4\alpha_5} = (\lm\gamma^m)_{\alpha_1}(\lm\gamma^n)_{\alpha_2}(\lm\gamma^p)_{\alpha_3}(\gamma_{mnp})_{\alpha_4\alpha_5}\\[.3cm]
\bar T^{\alpha_1\alpha_2\alpha_3\alpha_4\alpha_5} = (\blm\gamma^m)^{\alpha_1}(\blm\gamma^n)^{\alpha_2}(\blm\gamma^p)^{\alpha_3}(\gamma_{mnp})^{\alpha_4\alpha_5}
\ee
and they satisfy
\be
T_{\alpha_1\alpha_2\alpha_3\alpha_4\alpha_5} \bar T^{\alpha_1\alpha_2\alpha_3\alpha_4\alpha_5} =5! 2^6 (\lambda\bar\lambda)^3
\ee
After integrating over the non zero modes, one is left with the integrations over the zero modes of $\lambda^\alpha,\bar\lambda_\beta, \theta^\gamma$ and $r_\delta$.  To make use of the pure spinor superspace identities which appear at tree level, one defines an arbitrary function of pure spinor variables $(\lambda,\bar\lambda,r,\theta)$ as \cite{Gomez:2010ad}
\be
\Bigl\la M(\lambda,\bar\lambda,\theta,r)\Bigl\ra_{(n,g)}\equiv \int [d\lambda][d\bar\lambda][dr]  [d\theta] \ \f{e^{-\lm\blm-r\theta }}{(\lambda\bar\lambda)^{3-n}}\ M(\lambda,\bar\lambda,\theta,r)\label{A.67braket}
\ee
 The normalization for this bracket is fixed by setting $M=(\lambda^3\theta^5)$ which gives 
\be
\bigl\la (\lambda^3\theta^5)\bigl\ra_{(n,g)}&=&{N}_{n,g}\bigl\la (\lambda^3\theta^5)\bigl\ra\qquad,\qquad N_{(n,g)}\equiv 2^7R\left(\f{2\pi}{A_g}\right)^{5/2}\left(\f{\alpha'}{2}\right)^{2} \f{(7+n)!}{7!} \label{normpg}
\ee
where $A_g$ is the area of the Riemann surface and $R$ is the arbitrary constant which is fixed by the normalization of tree level amplitudes as mentioned above and the bracket $\la\cdots\ra$ can be evaluated by using the same pure spinor superspace identities which are used at tree level \cite{Berkovits_Mafra, Stanh}. 

\vspace*{.07in}The introduction of the bracket $\la\cdots\ra_{(n,g)}$ is quite useful because it satisfies the following identity \cite{Gomez:2013sla}
\be
\bigl\la  f(\lambda^{n+3},\bar\lambda^n,\theta) \bigl  \ra_{(n,g)}=\bigl\la (\lambda\bar\lambda)^n \hat f(\lambda^{3},\theta^5) \bigl  \ra_{(n,g)}\label{theorem1}
\ee
where, the functions $f$ can be expressed as 
\be
f(\lambda^{n+3},\bar\lambda^n,\theta)= \lambda^{\alpha_1}\cdots \lambda^{\alpha_{n+3}}\bar\lambda_{\beta_1}\cdots\bar\lambda_{\beta_n}f^{\beta_1\cdots\beta_n}_{\alpha_1\cdots\alpha_{n+3}}(\theta)
\ee
and $\hat f$ is given by
\be
\hat f(\lambda^3,\theta^5) = 672 \lambda^{\beta_1}\lambda^{\beta_2}\lambda^{\beta_3}\mathcal{T}^{\sigma_1\cdots\sigma_{n+3}}_{\beta_1\cdots\beta_{n+3}}f^{\beta_4\cdots\beta_{n+3}}_{\sigma_1\cdots\sigma_{n+3};\delta_1\cdots\delta_5}\theta^{\delta_1}\cdots\theta^{\delta_5}
\ee
where $\mathcal T$ is a symmetric and gamma traceless $SO(10)$ invariant tensor whose explicit form can be found in \cite{Gomez:2013sla}.

\subsection{First Massive States}
The heterotic closed string spectrum can be analyzed by considering the tensor product of the left and right moving states. For a detailed analysis of the first massive states, see e.g., \cite{witten}. At the first mass level, the supersymmetric sector has the same number of states as in the case of open strings, namely 128 bosonic and 128 fermionic states. The bosonic states represent a symmetric traceless second rank tensor with $44$ degrees of freedom and an anti-symmetric 3-form field with $84$ degrees of freedom. The 128 fermionic states represent a tensor spinor field. The Lorentz group transformation properties of these states can be represented as
\be
(\textbf{44},\textbf{1})\oplus(\textbf{84},\textbf{1})\oplus(\textbf{128},\textbf{1})
\ee
The first label corresponds to SO(9) and the second label corresponds to SO(32) group\footnote{More precisely, these are spin(9) and spin$(32)/Z_2$ \cite{witten}}.

\vspace*{.07in}The physical states are obtained by tensoring these with the states from the non supersymmetric sector. For the SO(32) heterotic string, the number of states in the non supersymmetric sector is 73764 which transform as
\be
(\textbf{44},\textbf{1})\oplus(\textbf{9},\textbf{496})\oplus(\textbf{1},\textbf{69256})
\ee
Again, the first and the second labels correspond to the transformation under the $SO(9)$ and $SO(32)$ groups respectively. These 73764 left moving states transform as scalars, spinors, 2nd rank anti-symmetric tensors, 4th rank anti-symmetric tensors and 2nd rank symmetric traceless tensors of SO(32).

\vspace*{.07in}Thus, the total physical degrees of freedom at the first massive level of SO(32) heterotic string is $256\times 73764= 18883584$. These states form the $N=1$ supermultiplet in 10 dimensions. A basic fact about these states is that they do not satisfy the BPS condition. However, they are still stable and undergo mass renormalization. 

\vspace*{.07in}As mentioned earlier, the supersymmetric sector of the world-sheet theory can be described by the pure spinor formalism. This means that the $256$ states in the supersymmetric sector can be described in terms of superfields in a manifestly supersymmetric invariant manner. Since these states are mathematically same as the ones which arise in the open string theory in 10 dimensions, we can use the results of \cite{Berkovits1,theta_exp} to describe these 256 states in terms of the superfields $B_{mnp}, G_{mn}$ and $\Psi_{m\alpha}$.

\section{ Some details about closed string vertex operators }
\label{appen:A}
In this appendix, we argue that the vertex operator for the closed strings can be obtained using the result for the open strings (see also \cite{Grassi:2004ih}). We consider the integrated and unintegrated vertex operators separately. 

\subsection{Unintegrated vertex operator}
The closed string unintegrated vertex operators satisfy 
\be
QV(z,\bar z)=0=\bar QV(z,\bar z)\qquad,\qquad \delta V = Q\Omega +\bar Q\bar\Omega\label{A.66gAuge}
\ee 
where, $\bar Q\Omega =0=Q\bar\Omega$. 

\vspace*{.07in}The above set of equations can be solved by assuming an ansatz of the form $V(z,\bar z)=V_R(z)V_L(\bar z)e^{ik\cdot X}$. With this ansatz, it is easy to see that the BRST equations can be satisfied provided 
\be
Q\left[V_R(z)e^{ik\cdot X}\right]=0\qquad,\quad \bar Q\left[V_L(\bar z)e^{ik\cdot X}\right]=0
\ee
Both the above equations are essentially the same equations which arise also in the open string case. We only need to take care of the difference in the OPEs of the open and closed strings while setting up these equations. Focussing on the right sector for simplicity, we recall that the main difference between the closed and open string expressions are in the definition of the cvariant derivative $D_\alpha$ and the OPE involving $\Pi^m$, namely
\be
D^{O}_{\alpha}=\p_\alpha +(\gamma_m\theta)_\alpha\p_m \quad &\rightarrow &\quad D^{C}_{\alpha}=\p_\alpha +\f{1}{2}(\gamma_m\theta)_\alpha\p_m \non\\[.3cm]
\Pi^O_m(z)V^O(w)=-\alpha'\f{\partial_m V^O(w)}{z-w}+\cdots\quad &\rightarrow &\quad \Pi^C_m(z)V^C(w)=-{\f{\alpha'}{2}}\f{\partial_m V^C(w)}{(z-w)}+\cdots\non
\ee
where the superfield $V$ include the factor $e^{ik\cdot X}$. 

\vspace*{.07in}We further note that if we use the expression of $\p_m$ in terms of $D_\alpha$, namely $\p_m= \f{1}{16}\gamma_m^{\alpha\beta}D^O_\alpha D^O_\beta$ for the open string and $\p_m= \f{1}{8}\gamma_m^{\alpha\beta}D^C_\alpha D^C_\beta$ for the closed strings, then the OPE expressions look exactly identical,
\be
\Pi^O_m(z)V^O(w)=-\f{\alpha'}{16}\gamma_m^{\alpha\beta}\f{D^O_\alpha D^O_\beta V^O(w)}{z-w}+\cdots\quad \rightarrow \quad \Pi^C_m(z)V^C(w)=-{\f{\alpha'}{16}}\gamma_m^{\alpha\beta}\f{D^C_\alpha D^C_\beta V^C(w)}{(z-w)}+\cdots\non
\ee
Since the BRST equation of motion only depend upon the OPEs, the above analysis shows that both the open and two sectors of the closed string equations take an identical covariant form when expressed using the covariant derivatives. 

\vspace*{.07in}Thus, an arbitrary BRST equation can be schematically written in the form 
\be
\sum_nf_n(D^O_\alpha)\left[\hat V^O_ne^{ik\cdot X}\right]=0\quad\rightarrow\quad \sum_nf_n(D^C_\alpha)\left[\hat V^C_ne^{ik\cdot X}\right]=0
\ee
In the above equation, $\hat V$ do not depend upon $X$ but only on the derivative of $X$. If we work in momentum space, the effect of $e^{ik\cdot X}$ is in replacing $\p_m$ in $D_\alpha$ by $k_m$. Hence, we can factor out $e^{ik\cdot X}$ in momentum space and can write 
\be
\sum_nf_n(D^O_\alpha)\left[\hat V^O_n\right]=0\quad\rightarrow\quad \sum_nf_n(D^C_\alpha)\left[\hat V^C_n\right]=0
\ee
If we write $D_\alpha$ as $D_\alpha = \p_\alpha +i\tilde k_m(\gamma_m\theta)_\alpha$, where $\tilde k=k^O$ for open string and $k=\f{1}{2}k^C$ for the closed string, then it is clear that the functional form of the superfield $\hat V$ for both open and closed strings will be same and we can obtain the left and/or right sector of the closed string vertex operators by replacing the momenta $k$ by $\f{1}{2}k$ in the open string vertex operator
\be 
\hat{V}^{C}=\hat{V}^{O}\left(k\rightarrow \f{1}{2}k\right)
\ee
and hence, the full pure spinor closed string vertex operator can be expressed as 
\be
V(k,z,\bar z)= V_R\left(\f{k}{2},z\right)V_L\left(\f{k}{2},\bar z\right)e^{ik\cdot X}
\ee
(where both $V_L$ and $V_R$ have the same functional form as in open string case) for type II and 
\be
V(k,z,\bar z)= V_R\left(\f{k}{2},z\right)V_L\left(\bar z\right)e^{ik\cdot X}
\ee
(where $V_R$ has same functional form as open string case and $V_L$ is constructed using SO(32) or $E_8\times E_8$ fields as in RNS) for the heterotic case.

\vspace*{.07in}In the above analysis, we have not yet taken into account the gauge freedom given in \eqref{A.66gAuge}. However, both the gauge superfields $\Omega$ and $\bar\Omega$ in the case of type II theories (or just $\Omega$ in the case of heterotic theories) have the same number of degrees of freedom as the single gauge superfield of open strings. Hence, once we obtain the full closed string vertex operators, they can be gauge fixed in exactly the same manner as in the open string case. See \cite{Berkovits:2004px} for this analysis  in explicit detail for the massless states.

\subsection{Integrated vertex operator}
The integrated vertex operator satisfies
\be
 Q\bar Q U(z,\bar z) = \p\bar\p V(z,\bar z)\label{1.2.14}
\ee
Using the form of $V$ described above, the right hand side of \eqref{1.2.14} can be written as
\be
\p\bar\p V &=& \p\bar\p \Bigl(V_L(z)V_R(\bar z)e^{ik\cdot X(z,\bar z)}\Bigl)\non\\
&=&\p\Bigl( V_L\bar\p V_R e^{ik\cdot X}+V_LV_R\bar\p e^{ik\cdot X}      \Bigl)\non\\
&=&\p V_L\bar\p V_R e^{ik\cdot X}+ V_L\bar\p V_R \p e^{ik\cdot X}+\p V_L V_R \bar\p e^{ik\cdot X}+ V_L V_R \p\bar\p e^{ik\cdot X}\label{1.2.16}
\ee
To satisfy \eqref{1.2.14}, we propose
\be
U(z,\bar z)= U_L(z)U_R(\bar z)e^{ik\cdot X}\label{1.2.17}
\ee
such that
\be
Q\Bigl(U_Le^{ik\cdot X}\Bigl)= \p\bigl(V_L e^{ik\cdot X}\bigl)\quad,\qquad \bar Q\Bigl(U_Re^{ik\cdot X}\Bigl)= \bar\p\bigl(V_R e^{ik\cdot X}\bigl)\label{1.2.18}
\ee
With this, the left hand side of \eqref{1.2.14} becomes
\be
Q\bar Q U &=& Q\Bigl[U_L\bar Q\bigl(U_R e^{ik\cdot X}\bigl)\Bigl]\non\\
&=&Q\Bigl[U_L\bar\p V_R e^{ik\cdot X}+U_L V_R \bar\p e^{ik\cdot X} \Bigl]\non\\
&=&Q\bigl(U_Le^{ik\cdot X}\bigl)\bar\p V_R + Q\bigl(U_L\bar\p  e^{ik\cdot X} \bigl)V_R\non\\
&=&\p\bigl(V_L e^{ik\cdot X}\bigl)\bar\p V_R + \bar\p \p\bigl(V_L e^{ik\cdot X}\bigl)V_R\non\\
&=&\p V_L\bar\p V_R e^{ik\cdot X}+ V_L\bar\p V_R \p e^{ik\cdot X}+\p V_L V_R \bar\p e^{ik\cdot X}+ V_L V_R \p\bar\p e^{ik\cdot X}
\ee
This is same as the left hand side \eqref{1.2.16}. This shows that the ansatz \eqref{1.2.17} satisfies the BRST equation \eqref{1.2.14} provided \eqref{1.2.18} holds. 

\vspace*{.07in}Thus, the construction of the integrated vertex has reduced to solving the two decoupled equations given in \eqref{1.2.18}. These equations also depend upon the OPEs of the theory and have the same form as the corresponding equations in open string case. Hence, using the same arguments as described above for the unintegrated vertices, we see that the $U_L$ and $U_R$ for the type II theories (or $U_R$ for the heterotic theories) have exactly the same functional form as the open string integrated vertex but with momenta $k^m$ replaced by $\f{1}{2}k^m$.

\subsection{Theta Expansion}
\label{theta_exp}
From the difference between the open and closed strings BRST equations of motion satisfied by the superfields, it follows that the theta expansion of the basic superfields for the closed strings can be obtained from the corresponding open string theta expansion by replacing $k^m\rightarrow \f{1}{2}k^m$. Thus, using the open string theta expansion results of \cite{theta_exp}, we obtain for the closed strings
\be
&&\Psi_{s\beta}\non\\
&=&\psi_{s\beta}+\f{1}{16}(\gamma^m\theta)_\beta\ g_{sm}-{\f{i}{48}}(\gamma^{mnp}\theta)_\beta k_{m}b_{nps}-{\f{i}{288}}(\gamma_{s}^{\;\;npqr}\theta)_{\beta}k_{n}b_{pqr} \non\\
&&-{\f{i}{4}}k^p(\gamma^m\theta)_{\beta}(\psi_{(m}\gamma_{s)p}\theta)-{\f{i}{8}}k_{m}(\gamma^{mnp}\theta)_{\beta}(\psi_{[s}\gamma_{np]}\theta)-{\f{i}{48}}(\gamma_{s}^{\;\;mnpq}\theta)_{\beta}k_{m}(\psi_{q}\gamma_{np}\theta)\non\\
&&-{\f{i}{48}}\alpha'k_mk^rk_s(\gamma^{mnp}\theta)_{\beta}
(\psi_p\gamma_{rn}\theta)+{\f{i}{2304}}\alpha'(\gamma^{mnp}\theta)_\beta k_mk^rk_s(\theta\gamma^q_{\;\;nr}\theta)\ g_{pq}\non\\
&&-{\f{i}{384}}(\gamma^{mnp}\theta)_\beta k_m(\theta\gamma^q_{\;\;[np}\theta) g_{s]q}-{\f{i}{2304}}(\gamma_{smnpq}\theta)_{\beta}k^{m}(\theta\gamma_{npt}\theta)\ g^{qt}\non\\
&&-{\f{i}{192}} k^p(\gamma^m\theta)_{\beta}(\theta\gamma_{pq(s}\theta)\ g_{m)q}-{\f{1}{6912}}(\gamma^{mnp}\theta)_\beta k_m(\theta\gamma^{tuvw}_{\;\;\;\;\;\;\;\;\;nps}\theta)k_tb_{uvw}\non\\
&&-{\f{1}{864\alpha'}}(\gamma_s\theta)_{\beta}(\theta\gamma^{npq}\theta)b_{npq}-{\f{1}{41472}}(\gamma_{s}^{\;\;mnpq}\theta)_{\beta}k_{m}(\theta\gamma_{tuvwnpq}\theta)k^tb^{uvw}\non\\
&&-{\f{1}{3456}}(\gamma^m\theta)_{\beta}(\theta\gamma^{npq}\theta)b_{npq}k_{m}k_s-{\f{1}{2304}}(\gamma_{smnpq}\theta)_{\beta}k^{m}(\theta\gamma^{tun}\theta)b_u^{\;\;pq}k_t\non\\
&&-\f{1}{96\alpha'}(\gamma^m\theta)_{\beta}(\theta\gamma^{qr}_{\;\;\;\;(s}\theta)b_{m)rq}+{\f{1}{384}}(\gamma^m\theta)_{\beta}(\theta\gamma^{nqr}\theta)k_nk_{(s}b_{m)qr}\non\\
&&+{\f{1}{384}}(\gamma^{mnp}\theta)_\beta k_m(\theta\gamma^r_{\;\;q[n}\theta)b_{ps]r} k^{q} +O(\theta^4)
\ee
and
\be
B_{\alpha\beta}&=&\gamma^{mnp}_{\alpha\beta}\Biggl[b_{mnp}+12(\psi_p\gamma_{mn}\theta)+{6}\alpha'k^rk_{m}(\psi_p\gamma_{rn}\theta)+\f{3}{8}(\theta\gamma_{mn}^{\;\;\;\;\;q}\theta)\ g_{pq}-{\f{3i}{8}}(\theta\gamma^{tu}_{\;\;\;\;m}\theta)k_{t}b_{unp}\non\\
&&\hspace*{.4in}+{\f{3}{16}}\alpha'k^rk_{m}(\theta\gamma_{rn}^{\;\;\;\;\;q}\theta)\ g_{pq}
-{\f{i}{48}}(\theta\gamma_{tuvwmnp}\theta)k^{t}b^{uvw}-{\frac{1}{12}} i k_{{s}} \left(\psi _{{v}} \gamma _{{t} {u}} \theta \right) \left(\theta  \gamma _{stuv m n p} \theta \right)\non\\
&&\hspace*{.4in}-{\f{1}{2}} i \alpha  k_{{s}} k_{{t}} k_m \left(\theta  \gamma _{{t} {u} n} \theta \right) \left(\psi _p \gamma _{{s} {u}} \theta \right)+{\f{i}{2}} k_{{s}} \left(\theta  \gamma _{{t} m n} \theta \right) \left(\psi _p \gamma _{{s} {t}} \theta \right)+{\f{i}{2}} k_{{s}} \left(\theta  \gamma _{{t} m n} \theta \right) \left(\psi _{{t}} \gamma _{{s} p} \theta \right)\non\\
&&\hspace*{.4in}+ {i} k_{{s}} \left(\theta  \gamma _{{s} {t} m} \theta \right) \left(\psi _n \gamma _{{t} p} \theta \right)-{\f{i}{2}} k_{{s}} \left(\theta  \gamma _{{s} {t} m} \theta \right) \left(\psi _{{t}} \gamma _{n p} \theta\right)+\frac{1}{64 \alpha'} {(\theta \gamma_{s m n} \theta)}
{(\theta \gamma_{t u p}  \theta)}b_{s t u}\non\\
&&\hspace*{.4in}
-\frac{1}{288 \alpha'} {(\theta \gamma_{s t u} \theta)}{(\theta \gamma_{m n p}  \theta)}b_{s t u} +\frac{1}{64 \alpha'}
{(\theta \gamma_{s t u} \theta)}{(\theta \gamma_{u n p} \theta)} b_{s t m} \non\\
&&\hspace*{.4in}+{\frac{1}{128} }{(\theta \gamma_{s u x} \theta)}{(\theta \gamma_{t x p}  \theta)} b_{s m n } k_{t} k_{u}
-{\frac{1}{64}} {(\theta \gamma_{s u n} \theta)}{(\theta \gamma_{t x p}   \theta)} b_{s t m}k_{u} k_{x}\non\\
&&\hspace*{.4in}+{\frac{1}{256}} {(\theta \gamma_{s t x} \theta)}{(\theta \gamma_{u n p}   \theta)}b_{s t m }k_{u} k_{x}
+{\frac{1}{768}} {(\theta \gamma_{x z m} \theta)}{(\theta \gamma_{s t u y z n p}  \theta)}  b_{s t u}k_{x} k_{y}\non\\
&&\hspace*{.4in}+{\frac{1}{768}} {(\theta \gamma_{u y z} \theta)}{(\theta \gamma_{s t x z m n p}   \theta)}b_{s t u} k_{x} k_{y}
+{\frac{1}{13824}} {(\theta \gamma_{s t u w x y z} \theta)}{(\theta \gamma_{v x y z m n p}   \theta)} b_{s t u}k_{v} k_{w}\non\\
&&\hspace*{.4in}+{\frac{1}{128}} {(\theta \gamma_{s v n} \theta)}{(\theta \gamma_{t u p}  \theta)}b_{s t u}  k_{v} k_{m}
+{\frac{1}{256}} {(\theta \gamma_{t u v} \theta)}{(\theta \gamma_{s n p}   \theta)} b_{s t u}k_{v} k_{m}\non\\
&&\hspace*{.4in}-{\frac{1}{384}} {(\theta \gamma_{s t u} \theta)}{(\theta \gamma_{v n p}   \theta)} b_{s t u}k_{v} k_{m}
-{\frac{1}{128}} {(\theta \gamma_{s t v} \theta)}{(\theta \gamma_{u v p}   \theta)}b_{s t m} k_{u} k_{n}\non\\
&&\hspace*{.4in}+{\frac{1}{768}} i  \left(\theta  \gamma _{{t} {v} {w}} \theta \right) \left(\theta  \gamma _{{s} {u} {v} {w} m n p} \theta \right)k_{{u}} g_{{s} {t}}+{\frac{1}{64}} i  \left(\theta  \gamma _{{s} {u} n} \theta \right) \left(\theta  \gamma _{{t} {u} p} \theta \right)k_{{t}} g_{{s} m}\non\\
&&\hspace*{.4in}+{\frac{1}{128}} i  \left(\theta  \gamma _{{s} {t} {u}} \theta \right) \left(\theta  \gamma _{{u} n p} \theta \right)k_{{t}} g_{{s} m}+{\frac{1}{128}} i \left(\theta  \gamma _{{s} m n} \theta \right) \left(\theta  \gamma _{{t} {u} p} \theta \right) k_{{u}} g_{{s} {t}}\non\\
&&\hspace*{.4in}+{\frac{1}{128}} i  \left(\theta  \gamma _{{s} {u} m} \theta \right) \left(\theta  \gamma _{{t} n p} \theta \right)k_{{u}} g_{{s} {t}}-{\frac{ i \alpha'}{128}}   \left(\theta  \gamma _{{s} {u} {v}} \theta \right) \left(\theta  \gamma _{{t} {v} p} \theta \right)k_{{t}} k_{{u}} k_n g_{{s} m}\; +\ O(\theta^5)\Biggl]
\ee
Using these theta expansion results, the theta expansion of the unintegrated as well as integrated vertex operators can be easily written down.

\section{ Some useful pure spinor identities}
\label{appen:PS}
The following identities turn out to be useful in simplifying the pure spinor correlators at the superfield level. For an arbitrary tensor-spinor superfield $\Psi_{m\alpha}$, we have
\begin{enumerate}
\item  $(\lambda \gamma _{mn}{\Psi}_{p})(\lambda {\gamma}_{m qrst} \lambda)=-(\lambda {\Psi}_{p})(\lambda {\gamma}_{nqrst} \lambda)$
\item $(\lambda {\gamma}_{mnpq} {\Psi}_{r}) (\lambda {\gamma}_{mnstu} \lambda)=-2(\lambda {\Psi}_{r}) (\lambda {\gamma}_{pqstu} \lambda)$
\item $(\lambda {\gamma}_{mnpqrs} {\Psi}_{t}) (\lambda {\gamma}_{mnpuv} \lambda)=6(\lambda {\Psi}_{t}) (\lambda {\gamma}_{qrsuv} \lambda)$
\item $(\lambda {\gamma}_{mnpqrstu} {\Psi}_{v}) (\lambda {\gamma}_{mnpqw} \lambda)=24(\lambda {\Psi}_{v}) (\lambda {\gamma}_{rstuw} \lambda)$
\end{enumerate}
These identities can be proved using the gamma matrix properties and the pure spinor constraint. E.g., to prove the first identity, we start by noting
\be
(\lambda \gamma _{mn}{\Psi}_{p})(\lambda {\gamma}_{m qrst} \lambda)&=&\bigl[\lambda (\gamma_m\gamma _{n}-\eta_{mn}){\Psi}_{p}\bigl](\lambda {\gamma}_{mqrst} \lambda)\non
\ee
The first term in the right hand side vanishes by pure spinor constraint. To see this, we use the identity $(\lambda\gamma^{abc}\lambda)=0$ to decompose the gamma 5-form term as $(\lambda {\gamma}_{mqrst} \lambda)=(\lambda \gamma_m{\gamma}_{qrst} \lambda)$. With this, the first term of the right hand side vanishes by the identity $(\lambda\gamma^m)_\alpha(\lambda\gamma_m)_\beta=0$ which follows from the pure spinor constraint. This proves the first identity. All the other identities can be proved in similar way. 

\vspace*{.07in}The following identities also turn out to be useful in simplifying the calculations after doing the theta expansion
\begin{enumerate}
\item $(\lambda\gamma^{mnpqr}\lambda)(\lambda\gamma_{mna}\theta)=0$
\item $(\lambda\gamma^{mnpqr}\lambda)(\lambda\gamma_{m}\theta)=0$
\item $(\lambda\gamma^{mnpqr}\lambda)(\lambda\gamma_{mnpab}\theta)=0$
\item $(\theta\gamma^{mnp}\theta)(\theta\gamma_{mnp}\theta)=0$
\end{enumerate}

These can also be proved using the pure spinor constraint by following the same method as described above.

\end{document}